# Evaluation of P3-type layered oxides as K-ion battery cathodes

Pawan Kumar Jha[†], Sanyam Nitin Totade[‡], Prabeer Barpanda[†#$] and Gopalakrishnan Sai Gautam[*‡]

[†]Faraday Materials Laboratory (FaMaL), Materials Research Centre, Indian Institute of Science, Bangalore 560012, India.

[‡]Department of Materials Engineering, Indian Institute of Science, Bengaluru, Karnataka 560012, India.

[#]Helmholtz Institute Ulm (HIU), Electrochemical Energy Storage, Ulm 89081, Germany.

[$]Institute of Nanotechnology, Karlsruhe Institute of Technology (KIT), Karlsruhe 76021, Germany.

[*]Corresponding Author; E-mail: **saigautamg@iisc.ac.in**

## Abstract

Given increasing energy storage demands and limited natural resources of Li, K-ion batteries (KIBs) could be promising next-generation systems having natural abundance, similar chemistry and energy density. Here, we have investigated the P3-type $K_{0.5}TMO_2$ (where TM = Ti, V, Cr, Mn, Co, or Ni) systems using density functional theory calculations, as potential positive intercalation electrodes (or cathodes) for KIBs. Specifically, we have identified the ground state configurations, and calculated the average topotactic voltages, electronic structures, on-site magnetic moments, and thermodynamic stabilities of all $P3-K_{0.5}TMO_2$ compositions and their corresponding depotassiated $P3-TMO_2$ frameworks. We find that K adopts the honeycomb or zig-zag configuration within each K-layer of all P3 structures considered, irrespective of the TM. In terms of voltages, we find the Co- and Ti-based compositions to exhibit the highest (4.59 V vs. K) and lowest (2.24 V) voltages, respectively, with the TM contributing to the redox behavior upon K (de-)intercalation. We observe all $P3-K_{0.5}TMO_2$ to be (meta)stable, and hence experimentally synthesizeable according to our 0 K convex hull calculations, while all depotassiated $P3-TMO_2$ configurations are unstable and may appear during electrochemical cycling. Also, we verified the stability of the prismatic coordination environment of K compared to octahedral coordination at the $K_{0.5}TMO_2$ compositions using Rouxel and cationic potential models. Finally, combining our voltage and stability calculations, we find $P3-K_xCoO_2$ to be the most promising cathode composition, while P3-$K_xNiO_2$ is worth exploring. Our work should contribute to the exploration of strategies and materials required to make practical KIBs.

## Introduction

Lithium-ion batteries (LIBs) have played a pre-eminent role in energy storage for the last three decades.[1-3] However, our excessive dependence on LIBs raises several challenges on natural abundance of critical elements, fragile supply chains, and cost,[4] which has motivated the search for intercalation chemistries that can be an alternative to LIBs. Notably, K-ion batteries (KIBs) have emerged as a viable alternative for economic grid-scale storage applications owing to the natural abundance of K, reversible (de)intercalation into graphite (as anode), and lower standard redox potential (K/$K^+$, -2.936 V vs. standard hydrogen electrode –SHE).[5-7] Despite several advantages over Li (and Na), the practical development of KIB is constrained by the need for robust cathode materials that can (de)intercalate $K^+$ reversibly. Several classes of materials have been explored as K-intercalation cathodes, namely layered oxides, polyanionic frameworks, Prussian blue analogous, and organic compounds.[5] Out of these, layered transition metal oxides (TMOs), similar to those used for Li and Na (de)intercalation, are promising in terms of their high theoretical energy density, high rate capability owing to the large two-dimensional $K^+$ diffusion pathways, and possible structural stability during cycling due to the large slab spacing that can suppress detrimental transition metal (TM) migration into K-layers.[8]

Typical K-containing layered structures, of composition $K_xTMO_2$ (x ≤ 1, TM = transition metal), exhibit a variety of stacking sequences, including prismatic-based P3 or P2, and octahedral-based O3 or O2, where the P3, P2, O3, and O2 notations are defined as per the nomenclature of Delmas and co-workers.[9] The type of coordination environment preferred by K (i.e., octahedral vs. prismatic, see **Figure S1** of the supporting information –SI, for an illustration) in a given framework, which in turn determines the type of stacking sequence of the structure, is primarily determined by the TM itself, the oxidation state(s) of the TM, and the K-concentration. For example, $KScO_2$ and $KCrO_2$ exhibit the O3 framework, while $K_{0.5}MnO_2$ adopts the P3 framework.[10-15] The relative stability of octahedral and prismatic coordination can also be quantified via the "cationic potential" model and/or the Rouxel diagram.[9,16] Notably, prismatic coordination is often stabilized at intermediate (x~0.5) or non-stoichiometric (x < 1) K-concentrations in layered frameworks, similar to observations in analogous Na-containing layered systems,[17-20] as illustrated by $K_xCrO_2$, $K_xMnO_2$, and $K_xCoO_2$ systems.[11,19] Indeed, Hagenmuller and co-workers' experiment-derived phase diagram of $A_xMO_2$ (A = Na or K; M = Cr, Mn, or Co), indicates that either the P3 or the P'3 is the stable phase at intermediate Na or K concentrations, irrespective of the TM.[11]

Experimentally, stoichiometric $KTMO_2$ (x=1) has been synthesized only in $K_xScO_2$, $K_xCrO_2$, $K_xFeO_2$, and $K_xMnO_2$ systems, with limitations on the observable electrochemical capacity.[10-15] As a result, previous studies have investigated non-stoichiometric $K_xTMO_2$ frameworks, often dealing with prismatic phases. For example, Vaalma $et$ $al.$ demonstrated $K_{0.3}MnO_2$ as a possible K-intercalation host,[21] which was followed by Kim $et$ $al.$'s report that found P3-$K_{0.5}MnO_2$ as a viable candidate as well.[22] Interestingly, the analogous P3-$Na_xMnO_2$ compound is metastable and cannot be trivially synthesized.[23,24] Hironaka $et$ $al.$ showed P3-$K_xCoO_2$ as an efficient reversible K-intercalation host,[25] while Hwang $et$ $al.$ developed P3-$K_{0.69}CrO_2$ $via$ electrochemical ion exchange from the parent O3-$NaCrO_2$ compound.[26] Notably, previous computational studies have revealed that the diffusivity of $K^+$ in prismatic stacking is higher than in octahedral stacking.[24,26,27] Thus, the existing literature indicates that K-containing layered TMOs with prismatic stacking sequences can be easily synthesized, exhibit reasonable cyclability, and good rate performance. However, systematic computational or experimental studies of P3-type K-containing layered TMOs are missing, so far.

Here, we have used density functional theory (DFT[28,29]) calculations to systematically evaluate various K-ion containing P3-layered oxides as candidate electrodes for KIBs. Specifically, we have calculated the lattice parameters, average intercalation voltage, thermodynamic stability, electronic properties, and on-site magnetic moments in $K_xTMO_2$ systems, where TM = Ti, V, Cr, Mn, Co, or Ni. We enumerate the possible in-plane K-ion orderings for $K_{0.5}TMO_2$ compositions, and determine the ground states using DFT. Subsequently, we evaluate the aforementioned properties for the ground state $K_{0.5}TMO_2$ configuration and its corresponding depotassiated composition, namely $TMO_2$. Notably, we observe that all P3-type $K_{0.5}TMO_2$ systems are thermodynamically stable, except V and Cr, with Co (Ti) system exhibiting the highest (lowest) predicted voltage of 4.59 V (2.24 V) vs. K. The ground state configurations for all $K_{0.5}TMO_2$ systems are identical, with K exhibiting a honeycomb or zig-zag ordering in each K-layer. Based on calculated projected density of states (pDOS) and on-site magnetic moments, we expect the TM to be redox-active upon K (de)intercalation in P3-$K_xTMO_2$. Also, we demonstrate the stability of prismatic over octahedral coordination in the $K_{0.5}TMO_2$ systems considered, via the Rouxel and cationic potential model approaches. Finally, based on voltage and stability metrics, we expect P3-

$K_xCoO_2$ and $K_xNiO_2$ to be promising candidates. We hope that our study will reinvigorate the computational and experimental investigations of P3-$K_xTMO_2$ systems as K-intercalating hosts.

## Computational Methods

We used the Vienna Ab Initio Simulation Package[30-32] to perform the DFT calculations, using the plane wave basis set with a kinetic energy cut-off of 520 eV and the projected-augmented wave (PAW[32, 33]) potentials to model the ionic cores, consistent with our previous work.[34,35] We sampled the irreducible Brillouin zone using Γ-centered Monkhorst-Pack[36] $k$-point meshes with a density of 32 points per Å, and we integrated the Fermi surface with a Gaussian smearing of width 0.05 eV. We relaxed the cell volume, shape, and ionic positions of all our structures without any symmetry constraints, till the atomic forces and the total energy were converged within |0.05| eV/Å and $10^{-5}$ eV, respectively.[37] All our calculations were spin-polarized, and we initialized the magnetic moments of all TM metals in a high-spin ferromagnetic ordering, except Co and Ni where we initialized with a low-spin ferromagnetic ordering for both the +3 and +4 oxidation states. For describing the electronic exchange-correlation, we utilized the Hubbard $U$ corrected, strongly constrained and appropriately normed (SCAN[38]) functional, i.e., SCAN+$U$. As derived in previous work, we used Hubbard $U$ corrections[39] of 2.5, 1.0, 2.7, 3.0, and 2.5 eV for Ti, V, Mn, Co, and Ni, respectively.[34,35] All pDOS calculations with the 'fake-self-consistent-field' procedure, as detailed in previous work.[40,41]

The starting structure for Ti, V, Mn, and Ni $K_{0.5}TMO_2$ compositions was the P3-$K_{0.3}MnO_2$, as obtained from the inorganic crystal structure database (ICSD[42]), where we created the Ti, V, and Ni structures via ionic substitution of Mn in P3-$K_{0.3}MnO_2$. We constructed the Cr- and Co-based structures based on previous reports.[10,25] Since several K-vacancy configurations are possible at the target $K_{0.5}TMO_2$ composition, we used the pymatgen package to enumerate all symmetrically distinct orderings[43] within each $K_xTMO_2$ supercell of size 2x2x1. We used VESTA for the visualization and illustration of structures used in our calculations.[44]

The redox reaction of topotactic (de-)intercalation of $K^+$ in a P3-type $TMO_2$ can be represented as:

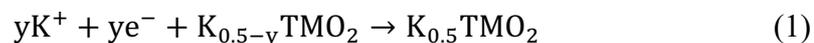

$$yK^+ + ye^- + K_{0.5-y}TMO_2 \rightarrow K_{0.5}TMO_2 \qquad (1)$$

$K_{0.5}TMO_2$ and $K_{0.5-y}TMO_2$ represent the potassiated and depotassiated structures, respectively. The average intercalation voltage can be calculated via the Nernst equation from the difference in the Gibbs energies ($G$) of the potassiated and depotassiated compositions. We approximated the Gibbs energies with the DFT calculated total energies (i.e., $G \approx E$), thus ignoring the $p - V$ and entropic contributions.[45,46] Within this approximation and with $F$ being the Faraday's constant, the average K-intercalation voltage *vs.* $K/K^+$ is:

$$V = \frac{-(E_{K_{0.5}TMO_2} - E_{K_{0.5-y}TMO_2} - yE_K)}{y.F} \qquad (2)$$

$E_{K_{0.5}TMO_2}$, $E_{K_{0.5-y}TMO_2}$, and $E_K$ are the DFT-calculated total energies of the ground state potassiated configuration, depotassiated composition, and the body-centered-cubic phase of pure K, respectively.

In order to assess the thermodynamic stability of the P3-type $K_{0.5}TMO_2$ and $TMO_2$ compositions, we computed the 0 K phase diagram (or the convex hull) of each ternary K-TM-O system, based on the DFT-calculated total energies of all elements, and compounds (i.e., binary and ternaries), whose experimentally reported structures are available in the ICSD. We used the pymatgen package to construct the phase diagrams.[43] We quantify the instability (stability) of $K_{0.5}TMO_2$ and $TMO_2$ by calculating the energy above (below) the hull, denoted by $E^{hull}$, based on the 0 K phase diagrams.[47-49] Note that we used a $E^{hull} \leq 50$ meV/atom as a threshold value for a structure being experimentally synthesizeable, but this threshold is arbitrary and is highly chemistry-dependent.[49]

## Results

### Structure, K-ordering, and lattice parameters

**Figure 1a** illustrates the typical unit cell of P3-$K_xTMO_2$, consisting of three $TMO_2$ layers, denoted by the brown polyhedra. The topologically distinct prismatic sites of K are shown by the blue and orange polyhedra in **Figure 1a** and equivalently by the blue and orange spheres in **Figure 1b**. Combined, the two prismatic sites arrange themselves in a hexagonal lattice, as shown by the black guidelines in panels b and c of **Figure 1**. Each $KO_6$ prism shares one of its triangular faces with one $TMO_6$ octahedra, thus violating the third Pauling's rule,[50,51] while the other triangular face shares its three edges with three different $TMO_6$ octahedra. The oxygen packing in P3 compounds follows the ABBCCA sequence.

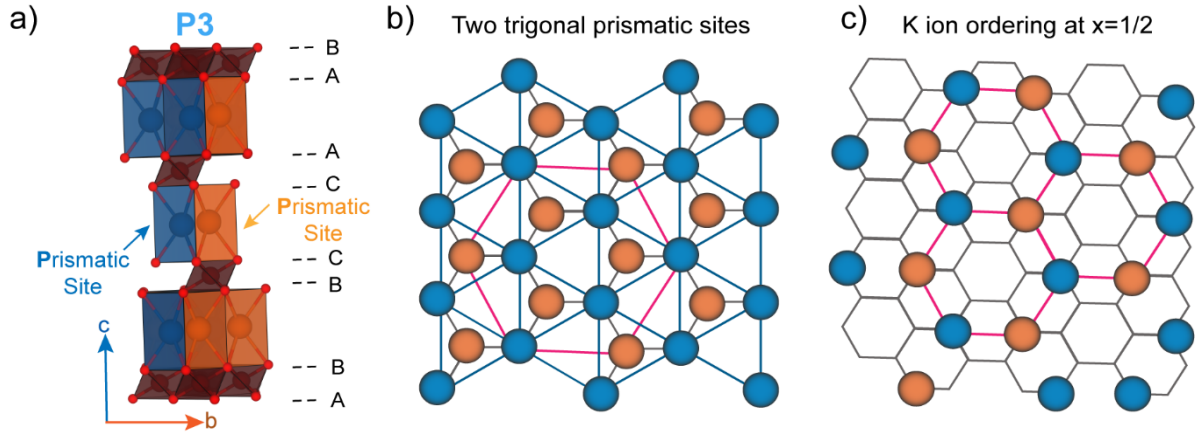

**Figure 1** (a) Unit cell of typical P3-$K_xTMO_2$ structure. (b) Visualization of different types of prismatic sites (blue and orange spheres) that are available for K-occupation, per K-layer in P3-$K_xTMO_2$. The blue and orange spheres are equivalent to the blue and orange polyhedra shown in panel a. (c) Honeycomb or zig-zag ordering in each K-layer that constitutes the ground state configuration for all P3-$K_{0.5}TMO_2$ considered. Pink guidelines in panels b and c connect identical K-sites.

The K-vacancy arrangement in each K-layer is an optimization of the steric and electrostatic interactions between the K-ions for any $K_{0.5}TMO_2$. Interestingly, we found the ground state K-vacancy configuration of all $K_xTMO_2$ systems to be identical, as indicated in **Figure 1c** by the honeycomb or zig-zag ordering of $K^+$ in each K-layer. Specifically, in each of the blue+orange hexagon of **Figure 1b**, $K^+$ occupies the farthest possible combination of one blue and one orange site, resulting in the honeycomb ordering of **Figure 1c**. This is equivalent to a K-K distance of $2a$ if the side-length of the hexagon in **Figure 1b** is $a$. Also, our in-plane K-ordering is similar to previous experimental and theoretical studies,[52-55] with marginal differences in the construction of the honeycomb ordering. Note that the scale of panels b and c in **Figure 1** are different, where the pink guidelines in both panels connect identical set of K-sites.

The SCAN+$U$-calculated lattice parameters for the ground state configurations of all $K_{0.5}TMO_2$ systems considered are compiled in **Table S1**. The computed lattice parameters are in good agreement with available experimental values,[22,25,26] with the maximum overestimation (underestimation) of the $c$ parameter of 4.69% (2.79%). Increase in the atomic number of the TM monotonically decreases the $a$ and $b$ lattice parameters (see trendlines in **Figure 2a**), caused primarily by a decrease in TM-O bonds, except for Mn and Ni which can be attributed to the Jahn-Teller distortion of $Mn^{3+}$ and $Ni^{3+}$ **(Figure S2)**. As shown in **Figure 2a** and **Table S1**, the Jahn-Teller distortion also causes a decrease in the $c$ parameter for Mn compared to Cr and Co, whereas in Ni, the $c$ parameter remains similar to Co.

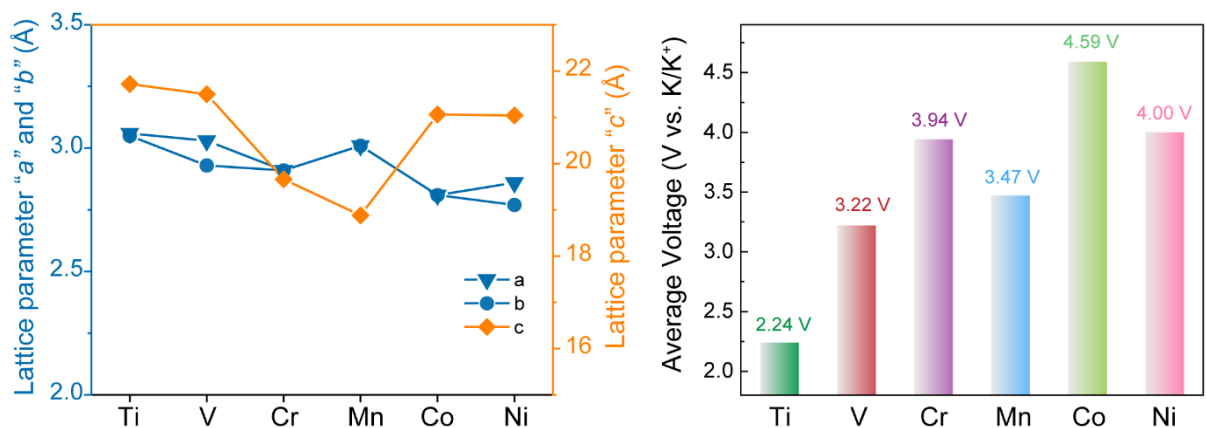

**Figure 2** SCAN+$U$ calculated (a) lattice parameters in various $K_{0.5}TMO_2$ systems, and (b) average K-intercalation voltage across the different $K_{0.5}TMO_2$-$TMO_2$ compositions. Average voltage values in (b) are also indicated by the text annotations.

## Average Voltages

**Figure 2b** presents the calculated average topotactic intercalation voltage, referenced against $K/K^+$, for the P3-type $K_{0.5}TMO_2$-$TMO_2$ systems considered in this work. Notably, the predicted voltages range from 2.24 V for the $K_{0.5}TiO_2$-$TiO_2$ system to 4.59 V for the $K_{0.5}CoO_2$-$CoO_2$ system, which is within the stable window of typical electrolytes used for KIBs.[5] The calculated intercalation voltage increases progressively as we move from Ti to Co, in accordance with the standard reduction potentials of the corresponding TMs, as noted in a previous study.[40] Importantly, Mn and Ni systems show markedly lower voltages compared to their neighboring TMs, which can be partly attributed to the Jahn-Teller distortions of $Mn^{3+}$ and $Ni^{3+}$. Also, the predicted voltage drop from Co to Ni is similar to the observation in Li-containing layered oxides, caused by the filling of the antibonding $e_g$ orbital of $NiO_2$.[40] Finally, the lower intercalation voltage of 2.24 V for the $K_xTiO_2$ system suggest that this system can be explored as an anode for KIBs.

**Electronic Structure and Magnetic Moments**

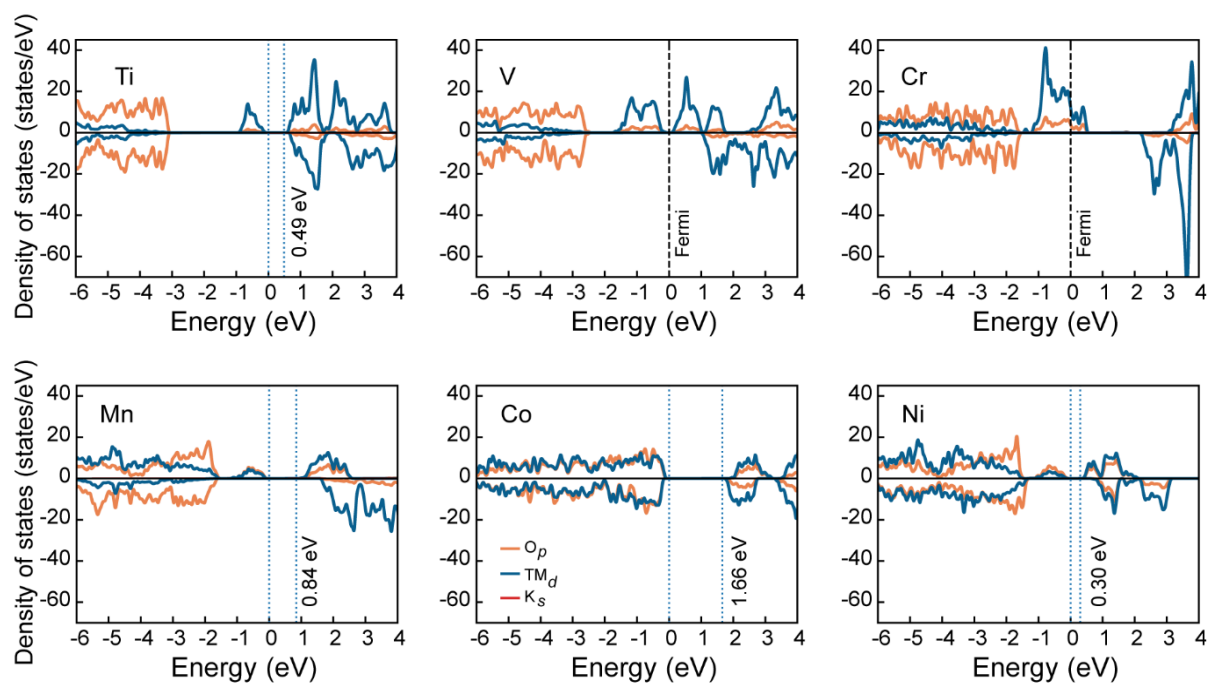

**Figure 3** SCAN+$U$-calculated pDOS for all P3-K$_{0.5}$TMO$_2$ systems. Blue, orange, and red curves correspond to TM $d$, O $p$, and K $s$ states, respectively. Positive (negative) values of pDOS correspond to up (down) spin electrons. Dotted blue lines represent the valence and conduction band edges, with the numbers indicating band gap values. Dashed black lines signify Fermi level. The zero on the energy scale in each panel is referenced either to the valence band maximum or to the Fermi level.

The calculated pDOS of all K$_{0.5}$TMO$_2$ ground state structures are displayed in **Figure 3**, with **Figure S3** compiling the pDOS of all TMO$_2$ structures. The red, blue, orange, dotted blue, and dashed black lines represent K $s$ states, O $p$ states, TM $d$ states, band edges, and Fermi level, respectively, with the numbers in each panel indicating band gaps. Except K$_{0.5}$CrO$_2$, K$_{0.5}$VO2 and CrO$_2$, all K$_{0.5}$TMO$_2$ and TMO$_2$ structures are predicted to be semiconductors by SCAN+$U$. The calculated band gaps exhibit a non-monotonic trend as we move along the 3$d$ series in K$_{0.5}$TMO$_2$, decreasing from 0.49 eV in Ti to 0 eV in Cr, subsequently increasing up to 1.66 eV in Co and further decreasing to 0.30 eV in Ni. Band gap trends in TMO$_2$ structures (**Figure S3)** are similar to K$_{0.5}$TMO$_2$, with the gap decreasing from 2.85 eV to 0 eV from Ti to Cr, then increasing to 2.40 eV in Mn and finally decreasing to 1.39 eV in Ni.

While TM $d$ states dominate the valence band edge (VBE) or the Fermi level in Ti, V, and Cr versions of K$_{0.5}$TMO$_2$ (**Figure 3**), both O $p$ and TM $d$ states contribute equally in the case of Mn, Co, and Ni analogs, attributed to increased hybridization of the TM-O bonds as we move across the 3$d$ series. In the case of conduction band edges (CBEs), the TM $d$ states dominate from Ti to Mn, while O $p$ states contribute significantly alongside TM $d$ states in Co and Ni structures. In the case of depotassiated $TMO_2$ structures (**Figure S3**), O $p$ states dominate the VBE in $TiO_2$, $CoO_2$, and $NiO_2$, TM $d$ states dominate VBE in $VO_2$, while a mixture of O $p$ and TM $d$ states contribute to the VBE/Fermi level in $MnO_2$ and $CrO_2$. The CBE of $TMO_2$ structures are dominated by TM $d$ states, with the exception of $NiO_2$, where a mixture of O $p$ and Ni $d$ states contribute. Given that TM $d$ states contribute significantly to the VBE/Fermi level (responsible for oxidation) of all $K_{0.5}TMO_2$ and the CBE/Fermi level (responsible for reduction) of all $TMO_2$ structures (except $NiO_2$), we expect the TM to be predominantly redox-active during $K^+$ (de)intercalation across all P3 systems (with Ni being a possible exception).

To further probe the possible origins of redox activity in the P3 frameworks, we analysed the calculated on-site magnetic moments of the TM in each system, as tabulated in **Table S2**. In all ground state $K_{0.5}TMO_2$ configurations, we observe that half the TM ions are in 3+, and the rest in 4+ oxidation state, except $K_{0.5}CrO_2$, where the $d$ electrons appear delocalized across Cr centers due to its metallic nature. Upon K-removal, all the transition metals in all $TMO_2$ structures are in a uniform 4+ oxidation state, as suggested by the calculated magnetic moments (see **Table S2**), highlighting that the TM exclusively contributes to the redox activity with $K^+$ (de)intercalation. Also, we observe from the magnetic moments that each $K^+$ in $K_{0.5}TMO_2$ shares the triangular face with a $TM^{4+}$ octahedra and the triangular edges with three $TM^{3+}$ octahedra.

**Thermodynamic stability**

The $E^{hull}$ for the $K_{0.5}TMO_2$ and $TMO_2$ compositions are displayed as a heatmap in **Figure 4**, where blue (red) tiles indicate compositions that are stable (unstable). The solid green line across the legend bar in **Figure 4** signifies the 50 meV/atom stability threshold. The 0 K convex hulls of the K-TM-O ternaries (relevant for potassiated compositions) and TM-O binaries (relevant for depotassiated compositions) are compiled in **Figures S4** and **S5**, respectively. Importantly, the $E^{hull}$ data indicates high degree of stability for all $K_{0.5}TMO_2$ frameworks, with the exception of $K_{0.5}VO_2$ and $K_{0.5}CrO_2$, which are metastable with $E^{hull}$ of 47 and 13 meV/atom, respectively, below the 50 meV/atom threshold. Thus, we expect all P3-type potassiated compositions considered in this work to be experimentally synthesizable.

In case of depotassiated compositions, we find all P3-TMO$_2$ structures to be unstable, with $E^{hull}$ more than 50 meV/atom. Thus, we don't expect the synthesis of P3-TMO$_2$ configurations to be facile. However, during electrochemical cycling, the P3-TMO$_2$ structures may exist in a metastable manner, due to kinetic barriers to transform to the corresponding stable states. Notably, the lower extent of instability displayed by P3-CoO$_2$ and P3-MnO$_2$ ($E^{hull}$~83 meV/atom) is more promising than the other frameworks in terms of their ability to appear during electrochemical cycling and not decompose to other stable compositions. Finally, combining both stability and voltage metrics, we find P3-K$_x$CoO$_2$ to be the most promising cathode composition, while P3-K$_x$NiO$_2$ can also be explored as a candidate.

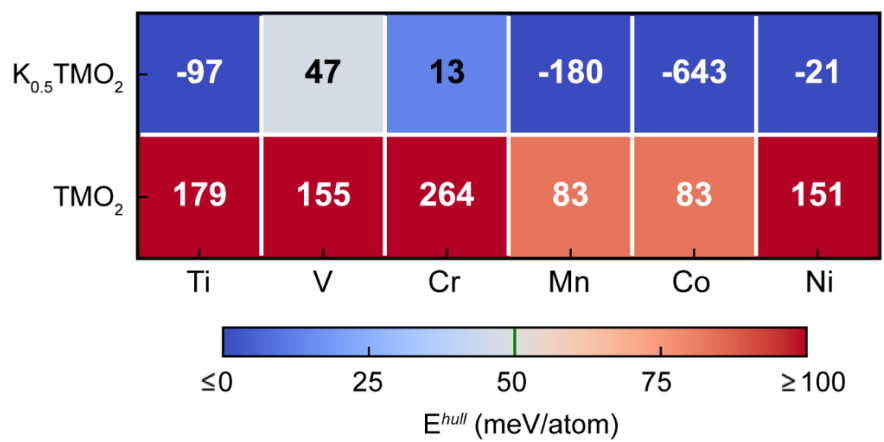

**Figure 4** DFT-calculated $E^{hull}$ for P3-K$_{0.5}$TMO$_2$ (top row) and P3-TMO$_2$ (bottom row) compositions. Each column represents a given TM. Blue (red) squares indicate high degrees of stability (instability), with the specific $E^{hull}$ value of each composition listed as a text annotation in the corresponding square. The green line on the legend bar indicates the rule-of-thumb $E^{hull}$~ 50 meV/atom threshold for experimental synthesizability.

## Discussion

Using DFT-based calculations, we have explored the P3-K$_{0.5}$TMO$_2$ frameworks as potential intercalation hosts for KIBs in this work. Specially, we have computed the lattice parameters, ground state K-vacancy configurations, average K-interaction voltage, electronic structure, on-site magnetic moments, and 0 K thermodynamic stability for P3-K$_{0.5}$TMO$_2$ and the corresponding depotassiated-TMO$_2$ structures, where TM is Ti, V, Cr, Mn, Co, or Ni. We found that all K$_{0.5}$TMO$_2$ ground states adopted the honeycomb or zig-zag ordering of K-ions. With respect to voltage predictions, we observed the highest (lowest) voltages to arise from the Co (Ti) system, consistent with trends in standard reduction potentials. While we found all potassiated P3-K$_{0.5}$TMO$_2$ compositions to be stable or metastable (i.e., $E^{hull} \leq 50$ meV/atom) highlighting experimental synthesizability, all depotassiated P3-TMO$_2$ compositions were unstable, indicating that they may not be synthesizable experimentally but may appear during electrochemical cycling due to kinetic barriers for decomposition. Also, we observed the TM to be the primary participant in the redox process, characterized by the electronic structure and on-site magnetic moments of the potassiated and depotassiated compositions. Finally, combining voltage and stability metrics, we find the P3 frameworks of K$_x$CoO$_2$, and K$_x$NiO$_2$ to be promising cathode candidates, while P3-K$_x$TiO$_2$ may be explored as an anode.

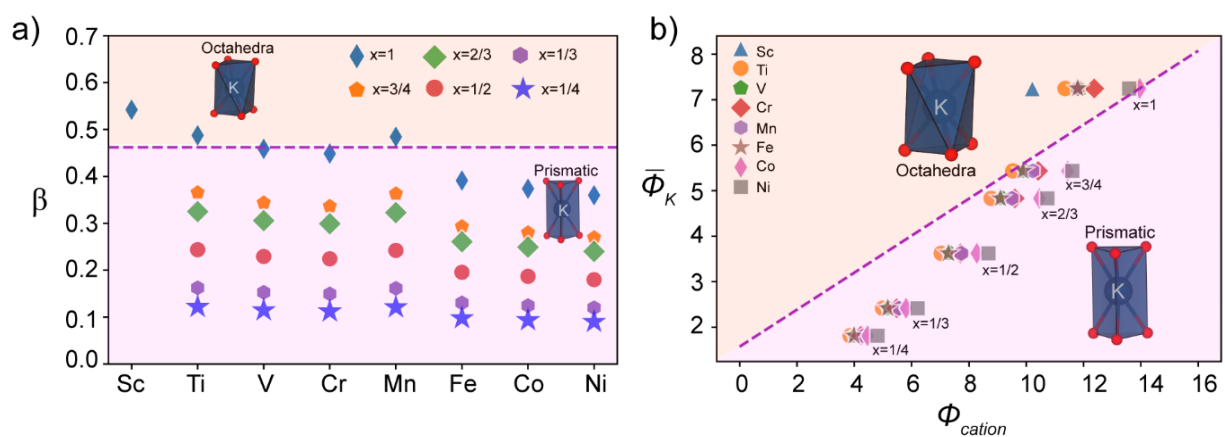

**Figure 5** (a) Rouxel diagram and (b) cationic potential phase map for various K$_x$TMO$_2$ compositions. The dashed line in both panels separates the region of stability of octahedral-coordinated phases from prismatic-coordinated phases. Each column of data points in panel a represents a distinct TM while the symbols represent various K compositions (x). In panel b, each row of data points corresponds to a unique x while the symbols distinguish the TMs.

Typically, in layered oxide frameworks, K can occupy either prismatic or octahedral coordination (**Figure S1**), depending on the K-concentration, and associated steric and electrostatic interactions within the structure. The relative stability of prismatic vs. octahedral coordination environment of K (and hence the stacking sequence of the layered structure) can be modelled via the modified Rouxel diagram[9] and the cationic potential[16] phase map, which are displayed in panels a and b of **Figure 5**, respectively. Numerical details of the Rouxel diagram and cationic potential frameworks are described in the SI, with **Tables S3-S6** compiling relevant parameters or values.

In the Rouxel diagram framework, the critical parameter (β) is used a classifier of prismatic and octahedral phases (see formulation provided in SI). β depends on the ionic or covalent nature of the bonds between the cations (K and TM) and anions (O), and the K concentration (x). **Figure 5a** plots β at different K compositions (x = 1/4, 1/3, 1/2, 2/3, 3/4, and 1) as the TM is varied in $K_xTMO_2$. Importantly, we observe that all $K_{0.5}TMO_2$ compositions prefer prismatic coordination, while several $KTMO_2$ compositions (except $KScO_2$, $KTiO_2$, and $KMnO_2$) also favour the prismatic coordination for any K content.

Cationic potential ($\phi_{cation}$) is a descriptor of interslab interactions, i.e., higher the cationic potential of a metal higher is the ionic polarizability and more covalent is the bond between the metal and an anion. Higher cationic potential also indicates stronger repulsion between adjacent $TMO_2$ octahedra (resulting from larger electrostatic repulsion of higher oxidation state metals) and weaker interaction between adjacent $KO_2$ slabs. The larger interlayer distance in a prismatic structure usually coincides with a higher cationic potential implying more covalent TM-O bonds. Conversely, the smaller interlayer distances in O3 structures (at similar compositions as corresponding P3 structures) coincide with a lower cationic potential.

Apart from interlayer distances, $\phi_{cation}$ is also dependent on K-concentration. For example, at low x in $K_xTMO_2$ (or equivalently lower mean ionic potential of K, $\bar{\phi}_K$), binding of $TMO_2$ layers by K via electrostatic attraction between $K^+$ and $O^{2-}$ is weaker resulting in larger interlayer spacings and prismatic stacking, which is what we observe in **Figure 5b**. Additionally, the variations in $\bar{\phi}_K$ and $\phi_{cation}$ for various $K_xTMO_2$ compositions indicate that prismatic structures should be observed at x = 1/2 for all TM, consistent with the Rouxel diagram framework as well (**Figure 5a**). Trends in panels a and b of **Figure 5** are also in agreement with the literature reported so far for the K-based layered oxides,[5] highlighting the utility of such empirical frameworks. Additionally, $\phi_{cation}$ can be tuned by the addition/doping of multiple TMs within the same layered oxide, thereby stabilizing either octahedral or prismatic coordination for the K-ions.

We find the honeycomb or zig-zag arrangement of $K^+$ to be the ground state configuration of all $K_{0.5}TMO_2$ compositions, as displayed in **Figure 1c**. Owing to the larger size and higher ionicity of $K^+$, there is strong in-plane electrostatic and steric repulsion that tends to maximize the $K^+$-$K^+$ distance, irrespective of the TM. Thus, $K^+$ can be considered to screen effectively TM-TM interactions across the layers (at x=0.5), which can be a cause of our observation of identical $K_{0.5}TMO_2$ ground states. Additionally, we find that each $K^+$ shares its triangular face and three triangular edges with $TM^{4+}$ and $TM^{3+}$, respectively, which may be a result of electrostatic interactions as well. Another consequence of the large size and higher ionicity of $K^+$ is the observed lower intercalation voltages than analogous Na-layered compounds, despite $K/K^+$ exhibiting a more negative standard reduction potential than $Na/Na^+$.[5, 40, 56] Specifically, the stronger electrostatic interactions between $K^+$ and the TM ions can result in increasing the interlayer distance, and could weaken the TM-O bonds by increasing the TM-O bond lengths compared to the Na-analogues.[57]

While we have used the SCAN+$U$ framework for describing the electronic exchange and correlation, recent studies have reported that SCAN+$U$ overestimates average voltages in Li-intercalation electrodes.[40, 41] Indeed, we observe a similar overestimation of average voltages in P3-$K_xCrO_2$, $K_xMnO_2$ and $K_xCoO_2$, where our predicted values are ~3.94 V, ~3.47 V and ~4.59 V vs. K, respectively (**Figure 2b**), compared to the experimental ~2.7 V in Cr (cathode composition was $K_{0.69}CrO_2$), ~2.7 V in Mn ($K_{0.5}MnO_2$), and ~ 3.1 V in Co ($K_{2/3}CoO_2$).[22, 25, 26] Such overestimation of voltages may arise from an underestimation of energies (i.e., total DFT-calculated energies are less negative) of metastable/unstable phases, such as depotassiated-$TMO_2$ structures.[40] Note that all our calculated voltages, despite being overestimated by SCAN+$U$, are within the stability window of the commonly used electrolyte, $KPF_6$ in ethylene carbonate: diethyl carbonate.

Finally, when it comes to phase stability, SCAN+$U$ frequently doesn't provide quantitative precision owing to underestimation of total energies of metastable phases. Yet, our calculations predicts that the traditional chemical synthesis pathway can produce P3-$K_{0.5}TMO_2$, owing to the calculated (meta)stability of P3-$K_{0.5}TMO_2$, whereas the depotassiated P3-$TMO_2$ may appear during electrochemical cycling. Apart from the single transition metal-based system, investigating possible inclusion of multiple TMs within the P3 framework could be result in better insertion host(s) for KIBs, where higher voltages arising from the presence of one TM can be combined with the stability contributed by another TM.

## Conclusion

We explored the K-containing P3-type layered TMOs as potential intercalation hosts for KIBs using DFT calculations and the SCAN+$U$ framework for describing electronic exchange and correlation. We considered six different TMOs, namely, $K_{0.5}TiO_2$, $K_{0.5}VO_2$, $K_{0.5}CrO_2$, $K_{0.5}MnO_2$, $K_{0.5}CoO_2$, and $K_{0.5}NiO_2$, and their corresponding depotassiated compositions, as the candidate K-intercalation hosts. Apart from estimating the ground state K-vacancy configuration in each TMO system considered, we evaluated the DFT-relaxed lattice parameters, topotactic average intercalation voltage, and 0 K thermodynamic stability. Additionally, we probed the nature of redox activity upon K (de)intercalation in these compounds by analysing the electronic structure and on-site TM magnetic moments in the potassiated and depotassiated structures. Importantly, we find that $K^+$ prefers the honeycomb or zig-zag ordering in $K_{0.5}TMO_2$, irrespective of the TM, highlighting the dominance of electrostatic interactions between $K^+$ ions within the same layer. Our calculated voltages follow the general trend of standard reduction potentials of the TMs involved, with the low-voltage P3-$K_xTiO_2$ framework being more suitable as an anode, and P3-$K_xCoO_2$ exhibiting the highest predicted voltage. Notably, we find the redox activity to be centered on the TM sites in all $K_xTMO_2$ systems, with negligible contribution from redox on the anionic sites. In terms of thermodynamic stability, we find all P3-$K_{0.5}TMO_2$ frameworks considered to be below the $E^{hull} = 50$ meV/atom threshold, indicating that synthesis of such compounds is likely to be facile. Finally, given the combination of our thermodynamic stability and average voltage estimates, we find P3-$K_xCoO_2$, and $K_xNiO_2$ as potential candidates as KIB cathodes.

# Conflicts of interest

There are no conflicts of interest to declare.

# Acknowledgments

G.S.G. acknowledges financial support from the Indian Institute of Science (IISc) Seed Grant, SG/MHRD/20/0020 and SR/MHRD/20/0013, and support from the Science and Engineering Research Board (SERB) of Government of India, under sanction numbers SRG/2021/000201 and IPA/2021/000007. P.B. is grateful to the Alexander von Humboldt Foundation (Bonn, Germany) for a 2022 Humboldt fellowship for experienced researchers. P.B. acknowledges financial support from the HP Green R&D Centre (Bangalore). P.K.J. and S.N.T. would like to thank the Ministry of Human Resource Development (MHRD), Government of India, for financial assistance. We also acknowledge the computational resources provided by the Supercomputer Education and Research Centre (SERC), IISc.